\documentstyle[preprint,prl,aps,eqsecnum]{revtex}

\def\inbar{\vrule height1.5ex width.4pt depth0pt}
\def\IC{\relax\hbox{\kern.25em$\inbar\kern-.3em{\rm C}$}}
\def\IP{\relax{\rm I\kern-.18em P}}
\def\IF{\relax{\rm I\kern-.18em F}}
\def\IZ{\relax\ifmmode\hbox{Z\kern-.4em Z}\else{Z\kern-.4em Z}\fi}
\def\IR{\relax{\rm I\kern-.18em R}}

\def\I1{\relax{\rm 1\kern-.35em 1}}

\newcommand\p{{\mbox{$\psi_1$}}} % p1 is not an allowed name
\newcommand\pp{{\mbox{$\psi_5$}}}
\newcommand\ppp{{\mbox{$\psi_R$}}}
\newcommand\pa{{\mbox{$\psi_\alpha$}}}
\newcommand\paa{{\mbox{$\psi_\beta$}}} % pb already exists?
\newcommand\NP{{\it Nucl. Phys.\ }}

\newcommand\PL{{\it Phys. Lett.\ }}
\newcommand\PR{{\it Phys. Rev.\ }}
\newcommand\PRL{{\it Phys. Rev. Lett.\ }}

\newcommand\Mod{{\it Mod. Phys. Lett.\ }}

\begin{document}
{\tighten 
\title{\bf Scattering of Several Multiply Charged Extremal $D=5$ Black Holes}
\author{David M. Kaplan and Jeremy Michelson}
\address{Department of Physics \\ University of California \\
	Santa Barbara, California  93106 \\
        e-mail: dmk,jeremy@cosmic1.physics.ucsb.edu}
\bigskip
\date{July 1, 1997}
\preprint{UCSBTH-97-15, hep-th/9707021}
\maketitle
\begin{abstract}
The moduli space metric for an arbitrary number of extremal $D=5$ black holes
with arbitrary relatively supersymmetric charges is found.
\end{abstract}

} % end the tighten

\section{Introduction} \label{intro}
One of the major advances in string theory over the last year and a half
has been the new understanding of the
properties of extremal, and some non-extremal, black holes
from a microscopic perspective.  This began with the microscopic D-brane
counting
of the entropy of some simple five-dimensional black holes, by
Strominger and Vafa~\cite{ascv}.  The entropy of innumerable other black
holes has since been counted in this way---for review and
references see e.g.~%
\cite{horrev,maldrev}.  Hawking radiation, including
grey-body factors,
has also been successfully derived from the microscopic
string theory~\cite{grey}.

In this paper, we will explore the scattering of certain five-dimensional 
extremal black holes that carry charges of more than one $U(1)$.
For most of
the paper, we will only discuss the classical, supergravity,
scattering; most of
the D-brane calculation is left for future work.  The standard approach
to the supergravity calculation follows the monopole scattering of
Manton~\cite{man}.  The idea is that, for extremal black holes, 
supersymmetry ensures that there
is no force between static black holes; hence any configuration of $N$~%
black holes has the same energy.  So, the positions of the black holes
are the $(D-1) N$ moduli of this sector of the theory (where $D$ is
the space-time dimension of the theory), and the slow
motion of the black holes is governed by the metric of the moduli space,
which is just the kinetic term of the low energy effective action.
The first such calculation was done for $D=4$ extremal
Reissner-Nordstr\"{o}m black
holes in~\cite{fe}.  Shiraishi~\cite{shir} then generalized this
to theories with a dilaton, in arbitrary dimension, and with arbitrary
coupling of the dilaton to the $U(1)$ vector field.  Khuri and Myers~%
\cite{myers2} then generalized this result to higher-dimensional
branes.  In particular, they found that for scattering of branes which
admit kappa-symmetric world-volume actions---and so, in particular, for
scattering of identical D-branes---there
is no interaction; the metric on the moduli
space is flat.

More recently, Douglas, Polchinski and Strominger \cite{dps} 
calculated the scattering
of D-brane probes off a $D=5$ extremal black hole that carried three
$U(1)$ charges.  They considered this
from both a supergravity and a D-brane point of view.
Their supergravity
calculation was performed by expanding the D-brane (Born-Infeld) action
in the background of the black hole.
The D-brane calculation involves reducing
the black-hole--D-string probe system to a 1+1-dimensional superconformal
effective Yang-Mills theory.
This calculation correctly
reproduced the leading-order supergravity
$\frac{q^2 v^2}{r^2}$ interaction, but not
the (same order in velocities but weaker) $\frac{q^3 v^2}{r^4}$
interaction.  This is still an unresolved problem.

With this motivation, we scatter these black holes
off each other, not probes.  Hence, this is again a Manton-type calculation.
In section~\ref{sugrasec} we discuss the string/supergravity theory
and the black hole solution.  In section~\ref{resultsec} we give
the calculation and the result.  Our result agrees, in the appropriate
limits, with the previous results of \cite{shir,myers2,dps}.%
\footnote{After this work was completed, ref.~\cite{gibbons} appeared
which discusses some further special cases.}
In section~\ref{bhcapsec} we discuss coalescence of two black holes. 
Section~\ref{discsec} concludes with some comments about D-brane approaches.

\section{Supergravity Solution}\label{sugrasec}

We will analyze scattering of the black holes of \cite{tseyt}.
The black holes arise in compactification of type IIB string theory on $T^5$.
All black holes in this
theory are U-dual to a black hole with at most three non-zero charges.
Black holes with less than three charges are singular at the horizon---the
dilaton (or another scalar) blows up there.  
Even for supersymmetric
multi-black hole solutions, one can always perform a U-duality
transformation so that only three $U(1)$s are turned on.
For
most of this paper we will, for simplicity, consider only black holes which
carry 10-dimensional NS-NS 5-brane charge and 1-brane (fundamental
string) charge, and momentum
along the 1-brane.  U-duality, and the fact that the cubic invariant of
$E_6$ is unique%
\footnote{It is inevitable that someone will worry that the duality group
is actually $E_6(\IZ)$, for which little is known, including uniqueness of
the cubic invariant.  However, type IIB supergravity is actually invariant
under the full $E_6$ duality, and furthermore, the classical
black hole solutions can be
continuously deformed to solutions with non-quantized charges.  Continuity
then requires that the moduli space metric be invariant under the full $E_6$.},
will then give the moduli space metric for generic (supersymmetric) black
hole configurations.  This will be discussed further below.
The ten-dimensional supergravity solution corresponding to these black holes
is easily obtained by an S-duality transformation of the corresponding RR-%
charged black holes of \cite{hms,jmphd,dps}.

Following the dimensional reduction
procedure of
\cite{maharana,cvetic}, the relevant bosonic terms in the 5-dimensional
Einstein action are found to be
\begin{eqnarray} \label{ssugra}
S = \frac{1}{16 \pi G}\int & d^5x \sqrt{-g} & \left\{ R-\frac{4}{3} 
          (\partial_\mu\varphi)^2
     + \frac{1}{8} \mbox{Tr}(\partial_\mu M L \partial^\mu M L) \right.
     \nonumber \\
& & \left. \mbox{} - \frac{1}{4} e^{\frac{8}{3} \varphi} \tilde{F}_{\mu \nu}
           \tilde{F}^{\mu \nu}
     - \frac{1}{4} e^{{-\frac{4}{3}} \varphi} {\cal F}^i_{\mu \nu} 
       (L M L)_{i j} {\cal F}^{j \mu \nu} \right\}.
\end{eqnarray}
Here, $\tilde{F}=d \tilde{A}$ is the $U(1)$ field strength
that is dual (in 5-dimensions) to the NS-NS
antisymmetric tensor field; an NS-NS 5-brane which wraps the $T^5$
appears to the 5-dimensional
observer to be carrying electric $\tilde{A}$ charge.  The $U(1)$ field
strengths ${\cal F}^i = d {\cal A}^i, i=\{1,\ldots,10\}$, are
Kaluza-Klein fields.  The first five come from $\hat{g}_{\mu I}$,
and the second five come from $\hat{B}_{\mu I}$ where $\hat{g}$ and
$\hat{B}$ are respectively the 10-dimensional metric and NS-NS
antisymmetric tensor, $\mu, \nu, \ldots = 0, \ldots, 4$
are $D=5$ space-time indices
and $I, J, \ldots = 5,\ldots,9$ label directions on the internal torus.
In particular, the 1-branes are chosen to wind
around the 5-direction, and hence carry
electric ${\cal A}^6$ charge, and the momentum on the string turns into
electric ${\cal A}^1$ charge.
The 5-dimensional string coupling constant is
$g = e^\varphi$ where $\varphi$ is the 5-dimensional dilaton.  The matrix of
scalars $M_{ij}$ is given by
\begin{equation} \label{defM}
M = \left( \begin{array}{cc} {\cal G}^{-1} & {-{\cal G}}^{-1} {\cal B} \\
     {-{\cal B}}^T {\cal G}^{-1} & {\cal G} + {\cal B}^T {\cal G}^{-1} {\cal B}
    \end{array} \right),
\end{equation}
where ${\cal G}_{IJ}$ and ${\cal B}_{IJ}$
are the internal components of the $D=10$ metric and antisymmetric tensor,
respectively, and
\begin{equation} \label{defL}
L = \left(\begin{array}{cc} 0 & \I1 \\ \I1 & 0 \end{array} \right).
\end{equation}
Finally, $G$ is the 5-dimensional Newton constant.

The static multi-black hole solution with the desired charges is given by (we
set $g=\alpha'=R=V=1$, where $R$ is the radius of the circle in the
5-direction and $V$ is the volume of the torus in the 6--9-directions)
\begin{mathletters} \label{soln}
\begin{eqnarray}
\label{metricsoln}
ds^2 &=& -\psi_1^{-\frac{2}{3}} \psi_5^{-\frac{2}{3}} 
          \psi_R^{-\frac{2}{3}} dt^2
    + \psi_1^{\frac{1}{3}} \psi_5^{\frac{1}{3}} \psi_R^{\frac{1}{3}}
      d\vec{x} \cdot d\vec{x}, \\
\label{a5soln}
\tilde{A} & =& \psi_5^{-1} dt, \\
\label{aRsoln}
{\cal A}^1 &=& \ppp^{-1} dt, \\
\label{a1soln}
{\cal A}^6 &=& \p^{-1} dt, \\
\label{phisoln}
e^\varphi &=& \pp^{\frac{1}{2}} \p^{-\frac{1}{4}} \ppp^{-\frac{1}{4}}, \\
\label{g55soln}
{\cal G}_{55} &=& \p^{-1} \ppp.
\end{eqnarray}
\end{mathletters}
All other fields are zero.
%Also, recall that $I,J,\ldots$ run from 5 to 9, so ${\cal G}_{55}$ is the
%top-left-most component.  
The $\pa$'s, ($\alpha = 1,5,R$)
are harmonic functions; for a solution with $N$ black holes,
\begin{equation} \label{defpsi}
\pa = 1 + \sum_{a=1}^N \frac{q_{\alpha a}}{r_a^2},
\end{equation}
where $q_{\alpha a}$ is the $\alpha$th charge of the $a$th black hole,
and $\vec{r}_a$ is its position, or more precisely, the location of the
horizon.  This coordinate patch does not cover the interior of the black
holes.  
This solution is BPS saturated and
generically preserves $\frac{1}{8}$th of the original $N=8$, 5-dimensional
supersymmetries.  If no
black hole carries one or two of the three types of charges,
then the preserved supersymmetry
increases to $\frac{1}{4}$ or $\frac{1}{2}$.  Saturation of the
Bogomol'nyi bound gives
the mass of the black holes as
\begin{equation}
m_a = q_{1a} + q_{5a} + q_{Ra}.
\end{equation}

The equations of motion for this solution require that the
Laplacian of each $\pa$ vanish.  This is true everywhere but at the
horizon, $\vec{x}=\vec{r}_a$.  There is no contradiction, however, as
the coordinate system is singular at the horizon; the equations of
motion hold everywhere the coordinates are good.  The equations of
motion will be satisfied in a coordinate
system which is non-singular at the horizon as well.

The reason we make this observation is that the black hole
scattering calculation requires the regularization of self-interaction
infinities.  This is achieved by smoothing out the black holes by
adding charged dust.  To find the correct couplings and normalization,
we add in sources at the horizons of the black holes in the equations
of motion.  Integrating the equations of motion gives 
\begin{eqnarray} \label{ssource}
S_{\mbox{\scriptsize source}} = -\frac{4 \pi^2}{16 \pi G}
    \sum_{a=1}^N \int & ds & \left\{
    - q_{aR} {\cal A}^1_\mu 
    \frac{dx_a^\mu}{ds} - q_{a1} {\cal A}^6_\mu \frac{dx_a^\mu}{ds}
    - q_{a5} \tilde{A}_\mu \frac{dx_a^\mu}{ds} 
    \right. \nonumber \\
&& \left. + e^{-\frac{4}{3}\varphi} q_{a5} + e^{\frac{2}{3} \varphi}
     {\cal G}^{-\frac{1}{2}}_{55} q_{aR}
    + e^{\frac{2}{3} \varphi} {\cal G}^{\frac{1}{2}}_{55} q_{a1} \right\}.
\end{eqnarray}
It must be emphasized that equation~(\ref{ssource}) is non-physical and
we include it only to smooth it out for regularization.
Smoothing out equation~(\ref{ssource}),
gives the (bosonic part) of the dust action
\begin{eqnarray} \label{sdust}
S_{\mbox{\scriptsize dust}} = 
  \frac{4 \pi^2}{16 \pi G} \int & d^5x \sqrt{-g} & \left\{
     \rho_R {\cal A}^1_\mu u^\mu + \rho_1 {\cal A}^6 u^\mu
     +\rho_5 {\tilde{A}}_\mu u^\mu \right. \nonumber \\
&& \left. - e^{-\frac{4}{3}\varphi} \rho_{5} - e^{\frac{2}{3} \varphi}
     {\cal G}^{-\frac{1}{2}}_{55} \rho_{R}
    + e^{\frac{2}{3} \varphi} {\cal G}^{\frac{1}{2}}_{55} \rho_{1} \right\}.
\end{eqnarray}
Here, $u^\mu = \frac{dx^\mu}{d\tau}$
is the 5-velocity of the dust.  
Also, we have suppressed
Lagrange multipliers and factors of $\sqrt{-u^\mu u_\mu} = 1$ that are
required to obtain the correct equations of motion from 
equation~(\ref{sdust}).

\section{Effective Action and Scattering} \label{resultsec}
The scattering calculation is very similar to that of~\cite{fe,shir},
but we will describe it in detail for completeness.  When the velocities
are small, the solution describing scattering black holes will be a
perturbation around the solution~(\ref{soln}) and~(\ref{defpsi}).
Furthermore, as there are no
linear terms in the action, solving equations of motion to
$O(\vec{v}=\frac{d\vec{x}}{dt})$
is sufficient, when substituting back into the action, to get the
effective action to $O(\vec{v}^2)$.  Then, by Galilean invariance,
the only perturbations of the fields are of the form
\begin{mathletters} \label{pert}
\begin{eqnarray}
\label{gpert}
\delta ds^2 = 2 \p^{-\frac{2}{3}} \pp^{-\frac{2}{3}}
    \ppp^{-\frac{2}{3}} \vec{Q}\cdot d\vec{x} dt, \\
\label{aRpert}
\delta \vec{{\cal A}^1} = \vec{P}_R - \ppp^{-1} \vec{Q}, \\
\label{a1pert}
\delta \vec{{\cal A}^6} = \vec{P}_1 - \p^{-1} \vec{Q}, \\
\label{a5pert}
\delta \vec{\tilde{A}} = \vec{P}_5 - \pp^{-1} \vec{Q},
\end{eqnarray}
\end{mathletters}
where $\vec{Q}$ and $\vec{P}_\alpha$ are first-order quantities to be
determined.

The $g_{ti}$ and $A_i$ equations of motion determine the exterior derivatives
of $\vec{Q}$ and $\vec{P}_\alpha$ to be
\begin{mathletters} \label{solved}
\begin{eqnarray}
\label{qsolved}
dQ = -\p \pp \ppp \sum_{\alpha} \pa^{-1} dK_\alpha, \\
\label{psolved}
dP_\alpha = -\p \pp \ppp \pa^{-1} \sum_{\beta \neq \alpha}
    \paa^{-1} dK_\beta,
\end{eqnarray}
where
\begin{equation} \label{defK}
\vec{K}_\alpha \equiv \pa \vec{v} = -4 \pi^2 \vec{\nabla}^{-2}
   (\p^{\frac{2}{3}}
   \pp^{\frac{2}{3}} \ppp^{\frac{2}{3}} \rho_\alpha \vec{v}).
\end{equation}
\end{mathletters}
Here we have used the equations of motion of the smoothed out
action $S+S_{\mbox{\scriptsize dust}}$, 
\begin{equation} \label{sourceeom}
\vec{\nabla}^2 \pa = -4 \pi^2 \p^{\frac{2}{3}}
   \pp^{\frac{2}{3}} \ppp^{\frac{2}{3}} \rho_\alpha.
\end{equation}
In principle, there should be functions of integration on the right
hand sides of equations~(\ref{qsolved}) and~(\ref{psolved}); however,
the exterior derivative of the equations results in homogenous
differential equations for the functions of integration, and they
can therefore be consistently set to zero.%

Substituting equations~(\ref{soln}) and~(\ref{pert})
into the action $S+S_{\mbox{\scriptsize dust}}$ of equations~%
(\ref{ssugra}) and~(\ref{sdust}) gives, after some tedious algebra
and integration by parts
\begin{eqnarray} \label{illbetoldtocutthisout}
S = \frac{1}{16 \pi G} \int & d^5 x & \left\{ \p \pp \ppp \left[
     -\sum_{\alpha < \beta}\pa^{-1}\paa^{-1} \partial_t \pa
         \partial_t \paa + 2 \pi^2 \p^{\frac{2}{3}}
          \pp^{\frac{2}{3}} \ppp^{\frac{2}{3}} \sum_\alpha \pa^{-1}
          \rho_\alpha \vec{v}^2 \right] \right. \nonumber \\
&&  + \left[ \sum_\alpha \partial_t \vec{P}_\alpha \cdot \vec{\nabla} \pa
       + 4 \pi^2 \sum_{\alpha} \p^{\frac{2}{3}} \pp^{\frac{2}{3}} 
         \ppp^{\frac{2}{3}} \rho_\alpha \vec{P}_\alpha \cdot \vec{v}
         \right] \nonumber \\
&&  \left. + \p^{-1} \pp^{-1} \ppp^{-1} \left[ -\frac{1}{2} |dQ|^2
      {-\frac{1}{4}} \sum_\alpha \pa^2 |dP_\alpha|^2
      + \frac{1}{2} \sum_\alpha \pa dP_\alpha \cdot dQ \right] \right\},
\end{eqnarray}
up to total derivatives.
We would like to write this entirely in terms of $\pa$, $\rho_\alpha$,
$\vec{v}^2$ and $\vec{K_\alpha}$.  This is done by using
equations~(\ref{solved}), and by noting that
time derivatives can be eliminated using $\partial_t = -\vec{v}\cdot \nabla$.
%The last square bracket is therefore trivial to evaluate.
%
%The middle square bracket of equation~(\ref{illbetoldtocutthisout})
%is a little trickier.  We can get the $P_\alpha$ to appear only in
%the combination $dP_\alpha$, as required, by substituting
%equation~(\ref{sourceeom}) in the second term, and then integrating
%by parts.  Then, the factor of
%$\vec{v}$ combines with the $\vec{\nabla} \pa$ to give $dK_\alpha$.
%Note, though, that the factor of $\vec{v}$ in the first term originated
%from a time derivative of $P_\alpha$; thus na\"{\i}vely, we would
%not associate the factor of velocity with the factor of $\pa$ in that
%term.  Rather, this is how the regularization occurs.
%
%This tells us what to do with the first term in the first square bracket
%of equation~(\ref{illbetoldtocutthisout})---it is (up to prefactors and
%a summation)
%$\partial_i \vec{K_\alpha} \cdot \vec{\nabla} K_{\alpha i}$.

Having done this, we take the black hole limit, which is
equation~(\ref{defpsi}) and
\begin{mathletters} \label{bhlimit}
\begin{eqnarray}
\label{denslimit}
\p^{\frac{2}{3}} \pp^{\frac{2}{3}} \ppp^{\frac{2}{3}} \rho_\alpha
    \vec{v}^2 & \rightarrow &  \sum_a q_{\alpha a} \delta^{(4)}(\vec{x}-
      \vec{r}_a) \vec{v}_a^2, \\
\label{Klimit}
\vec{K}_\alpha & \rightarrow &\sum_a \frac{q_{\alpha a}}{\vec{r}_a^2}
      \vec{v}_a^2,
\end{eqnarray}
\end{mathletters}
as clearly follows from equations~(\ref{sourceeom}) and~(\ref{defK}).
Integrating over space gives the final result:
\begin{eqnarray} \label{result}
S_{\mbox{\scriptsize eff}} = \frac{\pi}{4 G} \int & dt & \left\{
   -\sum_{a} m_a + \frac{1}{2} \sum_{a} m_a \vec{v}_a^2
   + \frac{1}{2} \sum_{a,b} (q_{1a} q_{5b} + q_{1a} q_{Rb} +
     q_{5a} q_{Rb}) \frac{|\vec{v}_a - \vec{v}_b|^2}{|\vec{r}_a-\vec{r}_b|^2} 
     \right. \nonumber \\
&& \left. + \frac{1}{4} \sum_{a,b,c} (q_{1a} q_{5b} q_{Rc} +
    q_{1a} q_{Rb} q_{5c} + q_{5a} q_{Rb} q_{1c}) |\vec{v}_a - \vec{v}_b|^2
    \right. \nonumber \\
&& \left. \times
    \left[\frac{1}{|\vec{r}_a-\vec{r}_b|^2 |\vec{r}_a-\vec{r}_c|^2}
       + \frac{1}{|\vec{r}_a-\vec{r}_b|^2 |\vec{r}_b-\vec{r}_c|^2}
       - \frac{1}{|\vec{r}_a-\vec{r}_c|^2 |\vec{r}_b-\vec{r}_c|^2} \right]
    \right\}.
\end{eqnarray}
Clearly, $a=b$ does not contribute in either sum.  Also, in the triple
sum, when $c=a$ or $b$, the divergent terms cancel leaving a finite
two-body interaction.  We also note that if two of the $N$ black holes
are coincident with identical velocities,
then equation~(\ref{result}) gives the result for
the corresponding configuration with $N-1$ black holes.

When $q_{5a} = q_{1a} = q_{Ra}$ for each $a$,
the
multi-black hole reduces to Shiraishi's $a=0$ black hole~\cite{shir}.
In this case, equation~(\ref{result}) agrees with the corresponding result
of~\cite{shir}.  
If two of the charges are zero---for example for scattering of 5-branes---%
then the scattering is trivial, as given in~\cite{myers2}.
If one of the charges is zero, then equation~(\ref{result})
reproduces the scattering of $(4\| 0)$-%
brane bound states of~\cite{tsch}, when the latter is reduced from
six to five dimensions.
Also, if we only consider scattering of two black holes, and let the
charges (and therefore the mass) of one of the black holes be much less than 
those of the other, 
the configuration reduces to that of~\cite{dps} and
equation~(\ref{result}) agrees with the results of that paper.

Equation~(\ref{result}) is easily made U-duality
invariant by
comparison to~\cite{dps}.  
After inserting factors
of $\alpha'$, $g=e^{\varphi_\infty}$, where the subscript denotes evaluation
at spatial infinity, and
\begin{equation} \label{planck}
l_p = \left(\frac{g^2 \alpha'^4}{VR}\right)^{\frac{1}{3}}
\end{equation}
where $R$ is the radius of the circle in the 5-direction, and $V$ is
the volume of the $T^4$ in the 6--9 directions, and in particular,
rescaling the charges appropriately (as was done carefully
in~\cite{dps}) the result is
\begin{eqnarray} \label{uresult}
S_{\mbox{\scriptsize eff}} = \int & dt & \left\{
   -\sum_{a} m_a + \frac{1}{2} \sum_{a} m_a \vec{v}_a^2
   + \frac{1}{2} \sum_{a<b} (m_a m_b l_p^3 - q_{\Lambda a} 
       ({\cal M}_\infty^{-1})^{\Lambda \Sigma} q_{\Sigma b})
     \frac{|\vec{v}_a - \vec{v}_b|^2}{|\vec{r}_a-\vec{r}_b|^2} 
     \right. \nonumber \\
&& \left. + \frac{1}{4} \sum_{a<b} \sum_{c} d^{\Lambda \Sigma \Gamma}
    q_{\Lambda a} q_{\Sigma b} q_{\Gamma c} l_p^3
    |\vec{v}_a - \vec{v}_b|^2 \right. \nonumber \\
&& \left. \times
    \left[\frac{1}{|\vec{r}_a-\vec{r}_b|^2 |\vec{r}_a-\vec{r}_c|^2}
       + \frac{1}{|\vec{r}_a-\vec{r}_b|^2 |\vec{r}_b-\vec{r}_c|^2}
       - \frac{1}{|\vec{r}_a-\vec{r}_c|^2 |\vec{r}_b-\vec{r}_c|^2} \right]
    \right\},
\end{eqnarray}
where $d^{\Lambda \Sigma \Gamma}$ is proportional to the cubic $E_6$
invariant, and ${\cal M}$ is the kinetic matrix for the vector fields in
the full, manifestly U-dual, 5-dimensional Einstein bosonic
supergravity action.

\section{Geodesic Motion and Black Hole Capture} \label{bhcapsec}
If we consider the case of two black holes scattering off each other,
the effective action reduces to
\begin{equation} \label{twobody}
S_{\mbox{\scriptsize 2b}} = \int  dt  \left\{
   -M + \frac{1}{2} M \vec{V}^2 + \frac{1}{2} f(r) \vec{v}^2 
	\right\},
\end{equation}
where $M=m_1 + m_2$, $V = (m_1 v_1 + m_2 v_2 )/M$, 
$r = |\vec{r}_1 - \vec{r}_2|$, $\vec{v} = \vec{v}_1 - \vec{v}_2$,
\begin{mathletters}\label{symbols}
\begin{eqnarray}
\label{function}
	f(r) & = & \left[ \mu + \frac{\Gamma_{II}}{r^2} +
		\frac{\Gamma_{III}}{r^4} \right] ,\\
\label{GammaII}
	\Gamma_{II} & = & (q_{1,1}q_{5,2}+q_{5,1}q_{R,2}+q_{R,1}q_{1,2})
		+ ( 1 \leftrightarrow 2), \\
\label{GammaIII}
	\Gamma_{III} & = & (q_{1,1}q_{5,2}q_{R,2} + q_{R,1}q_{1,2}q_{5,2}
		+ q_{5,1}q_{R,2}q_{1,2}) + (1 \leftrightarrow 2),
\end{eqnarray}
\end{mathletters}
and $\mu$ is the reduced mass.
This is the appropriate generalization of the results of~\cite{dps}.
Therefore, the metric on the two-body moduli space is (neglecting
center of mass motion)
\begin{equation} \label{2bdymetric}
ds^2_{\mbox{\scriptsize 2b}} = f(r) (dr^2 + r^2 d\Omega^2)
\end{equation}
As noted in~\cite{dps},
although this metric is euclidean, and
therefore has no horizons, generically there is a second asymptotic
region around $r=0$.  This can be seen by transforming to a new 
radial coordinate $\rho = 1/r$, in which the metric takes the form
\begin{equation} \label{trans2bdymetric}
ds^2_{\mbox{\scriptsize 2b}} = 
    \left[ \Gamma_{III} + \frac{\Gamma_{II}}{\rho^2} +
	\frac{\mu}{\rho^4} \right] (d\rho^2 + \rho^2 d\Omega^2).
\end{equation}

The analysis of geodesic motion in this moduli space proceeds in
exactly the same manner as in~\cite{dps}.
Given two black holes
approaching each other with asymptotic velocity $v_\infty$ and impact
parameter $b$, there is a critical impact parameter $b_c$ (independent
of $v_\infty$) below which the black holes must coalesce.  This is
given by  
\begin{equation} \label{bcoalesce}
\frac{b^2_c}{\mu} = \Gamma_{II} + 2 \sqrt{\mu \Gamma_{III}}.
\end{equation}
Black holes which scatter at this critical impact parameter will orbit
each other at radius
\begin{equation} \label{rcoalesce}
r^2_c = \sqrt{\Gamma_{III} / \mu}.
\end{equation}

%One may analyze the motion of two black holes approaching each other
%with asymptotic relative velocty $v$ and impact parameter $b$ by
%considering the conserved quantity
%\begin{equation} \label{energy}
%E = f(r)^{-1} \left[ \frac{p^2}{2} + \frac{L^2}{2r^2} \right],
%\end{equation}
%where $L=r^2 f(r) \dot{\theta}$ is the (conserved) angular momentum
%and $p = f(r) \dot{r}$ is the canonical radial momentum.
%Asymptotically, $E=\frac{\mu v^2}{2}$ and $L=bv$.  Solving for
%$\dot{r}$ gives
%\begin{equation}\label{rdot}
%\dot{r} = v f(r)^{-1} ( \mu f(r) - b^2/r^2)^{1/2}
%\end{equation}

\section{Discussion} \label{discsec}
We have derived the U-duality invariant metric on the multi-black hole
moduli space from the supergravity point of view.  It would be
interesting to derive this from a D-brane perspective.  We will
conclude this paper by discussing a couple of possible approaches.

One approach is a simple generalization of the D-brane probe
calculations of~\cite{dps}.
The main observation is that the calculation of~\cite{dps} involves
a $U(q_5)$, $1+1$-dimensional gauge theory, that is essentially a dimensionally
reduced 5-brane worldvolume theory, with extra fields due to the 1-branes
and their excitations.  There were also extra fields representing the D-string
probe and its interactions.  Considering instead multiple black holes
essentially only requires consideration of a $U(\prod_a q_{5a})$ gauge
theory, that is Higgsed to $U(q_{51}) \times \ldots \times U(q_{5N})$.
However, this calculation is (by construction!) so similar to the one
performed in~\cite{dps} that it will most likely similarly fail to
reproduce the cubic interaction.

An alternative approach is motivated by the observation that the cubic
interaction involves factors of $\frac{1}{r^2}$ that are suggestive of
strings stretched between D-branes.  Therefore, it is reasonable to
hypothesize that a string analysis of three spatially separated black
holes, each with a single non-trivial charge turned on, may be able
to reproduce the cubic interaction.  This analysis has not been done,
and is qualitatively different from that of~\cite{dps}, and therefore
is an interesting direction for future research.

\acknowledgments
We thank Gary Horowitz, Veronika Hubeny, Juan Maldacena, Rob Myers and
Joe Polchinski for discussions and suggestions.
We also thank Andy Strominger for
discussions, suggestions and an advance copy of~\cite{dps}.
We are also grateful to Harald H. Soleng for making~\cite{cartan}
available, which was useful in checking some of the computation.
J.M. thanks NSERC and NSF for financial support.
This work was supported in part by DOE Grant No. DOE-91ER40618.

\end{document}